\documentclass[prb,aps,preprint]{revtex4}
\usepackage{graphicx}

\begin{document}
\newcommand{\be}{\begin{equation}}
\newcommand{\ee}{\end{equation}}
\newcommand{\bea}{\begin{eqnarray}}
\newcommand{\eea}{\end{eqnarray}}

\title{Conformational space annealing and an off-lattice frustrated model protein}
\author{Seung-Yeon Kim$^1$, Sung Jong Lee$^2$,
and Jooyoung Lee$^1$\footnote{correspondence to: jlee@kias.re.kr}}
\affiliation{$^1$School of Computational Sciences,
Korea Institute for Advanced Study,
207-43 Cheongryangri-dong, Dongdaemun-gu, Seoul 130-722, Korea \\
$^2$Department of Physics and Center for Smart Bio-Materials,
The University of Suwon, Hwasung-si, Kyunggi-do 445-743, Korea}


\begin{abstract}
A global optimization method, conformational space annealing (CSA),
is applied to study a 46-residue protein with the sequence
$B_9N_3(LB)_4N_3B_9N_3(LB)_5L$, where $B$, $L$ and $N$ designate hydrophobic,
hydrophilic, and neutral residues, respectively.
The 46-residue BLN protein is folded into the native state of
a four-stranded $\beta$-barrel.
It has been a challenging problem to locate the global minimum of the 46-residue BLN
protein since the system is highly frustrated and consequently its energy landscape
is quite rugged.
The CSA successfully located the global minimum of the 46-mer for all 100
independent runs.
The CPU time for CSA is about seventy times less than that for simulated annealing (SA),
and its success rate (100\%) to find the global minimum is about eleven times higher.
The amount of computational efforts used for CSA is also about ten times less than
that of the best global optimization method yet applied to the 46-residue BLN protein,
the quantum thermal annealing with renormalization.
The 100 separate CSA runs produce the global minimum 100 times as well as
other 5950 final conformations corresponding to a total of 2361 distinct local minima
of the protein.
Most of the final conformations have relatively small RMSD values from the global
minimum, independent of their diverse energy values.
Very close to the global minimum, there exist quasi-global-minima which
are frequently obtained as one of the final answers from SA runs.
We find that there exist two largest energy gaps between the quasi-global-minima
and the other local minima.
Once a SA run is trapped in one of these quasi-global-minima,
it cannot be folded into the global minimum before crossing over
the two large energy barriers,
clearly demonstrating the reason for the poor success rate of SA.
\end{abstract}

\maketitle


\section{introduction}

Finding the global minimum of a given function,
called the global optimization, is an important problem in various
fields of science and engineering.
Many optimization problems are hard to solve
since many of them belong to the NP-complete class, where
the number of computing steps required to solve the problem increases
faster than any power of the size of the system.
Some of the well-known classic examples are
the traveling salesman problem in applied mathematics and computer science,
spin glasses in condensed-matter physics,
and the protein-folding problem in biophysics.

Many of global optimization methods have been developed
and successfully applied to a variety of problems.
Examples of such methods are
simulated annealing (SA) \cite{kirkpatrick,laarhoven},
genetic algorithm (GA) \cite{holland,goldberg},
Monte Carlo with minimization (MCM) \cite{li},
multicanonical annealing \cite{jlee94,xu},
and quantum thermal annealing \cite{ylee00a,ylee00b,ylee01}.
One of the simplest algorithms for unbiased global optimization is
the SA method which has been most widely used.
Although the SA is very versatile in that it can be easily applied practically to any
problem, the drawback is that its efficiency is usually much lower than
problem specific algorithms.
This is especially problematic for NP-complete problems.
For this reason, it is important to find an algorithm
which is as general as SA, and yet competitive with problem specific ones.
Recently, a powerful global optimization method called
conformational space annealing (CSA) \cite{jlee97,jlee98,jlee99a} was proposed,
and applied to protein stucture prediction \cite{jlee99b,liwo,jlee99c,jlee00}
and Lennard-Jones clusters \cite{julee}.
The benchmark tests \cite{jlee97,jlee98,jlee99a} have demonstrated
that it can not only find the known global-minimum conformations
with less computations than existing algorithms,
but also provide new global minima in some cases.

In this paper, we apply the CSA method to
an off-lattice model protein introduced by Honeycutt and Thirumalai
\cite{honeycutt90,honeycutt92,veitshans}.
The so-called BLN model proteins are constructed from the residues of three types,
hydrophobic ($B$), hydrophilic ($L$), and neutral ($N$) residues found in nature.
It is shown that they exhibit many similarities with real proteins.
With the aid of simulation methods Honeycutt and Thirumalai studied
a 46-residue BLN protein with the sequence $B_9N_3(LB)_4N_3B_9N_3(LB)_5L$
that folds into the native structure of a four-stranded $\beta$-barrel.
They suggested the metastability hypothesis that
a polypeptide chain may have a variety of metastable minima
corresponding to folded conformations
with similar structural characteristics but different energies.
This hypothesis suggests that the particular state into which a protein
folds depends on the initial condition of the environment,
and that there are multiple pathways for the folding process.
Thirumalai {\it et al.} \cite{guo92,thirumalai} also demonstrated
that the folding kinetics of the BLN 46-mer is very similar to that of real proteins.
It is shown \cite{guo95,guo97a,guo97b,shea} that the 46-mer has two characteristic
transition temperatures: the transition from a random coil state
to a collapsed but nonnative state and the transition from a collapsed state
to the native $\beta$-barrel state.
Subsequent studies \cite{berry,nymeyer,elmaci,miller,evans} illustrated that
the 46-mer system exhibits a high degree of frustration
and its energy landscape is very rugged.
That is, there are many local minima with energies close to that of the native state,
and it is very difficult to find the global minimum
\cite{xu,ylee00a,ylee01,amara}.


\section{BLN model protein}

The potential energy of a BLN protein with $M$ residues is given by
\be
V=V_b+V_a+V_t+V_v,
\ee
where $V_b$ is the bond-stretching energy,
$V_a$ the bond-angle bending energy, $V_t$ the torsional dihedral-angle energy,
and $V_v$ the van der Waals energy.
The bond-stretching energy is
\be
V_b=\sum_{i=1}^{M-1}{k_r\over2}(|\vec{r}_{i+1}-\vec{r}_i|-a)^2,
\ee
where the force constant is given by $k_r=400\epsilon/a^2$,
$\epsilon$ is the energy constant, $a$ the average bond length,
and $\vec{r}_i$ the position of the $i$-th residue.
The bond-angle bending energy is given by
\be
V_a=\sum_{i=1}^{M-2}{k_\theta\over2}(\theta_i-\theta_0)^2,
\ee
where the force constant is $k_\theta=20\epsilon/({\rm rad})^2$,
$\theta_i$ is the bond angle defined by three residues $i$, $i+1$ and $i+2$,
and $\theta_0=1.8326$ rad or $105^\circ$.
The torsional dihedral-angle energy is expressed as
\be
V_t=\sum_{i=1}^{M-3}[A_i(1+\cos\phi_i)+B_i(1+\cos3\phi_i)],
\ee
where $\phi_i$ is the dihedral angle defined by four residues $i$, $i+1$, $i+2$ and $i+3$.
The amplitudes $A_i$ and $B_i$ are given by $A_i=0$ and $B_i=0.2\epsilon$
if two or more of the four residues
are neutral. For all the other cases, $A_i=B_i=1.2\epsilon$.
Finally, the van der Waals energy is given by
\be
V_v=4\epsilon\sum_{i=1}^{M-3}\sum_{j=i+3}^M
C_{ij}\biggl[\Bigl({\sigma\over r_{ij}}\Bigr)^{12}
-D_{ij}\Bigl({\sigma\over r_{ij}}\Bigr)^6\biggr],
\ee
where $\sigma$ is the Lennard-Jones parameter
and $r_{ij}$ is the distance between non-bonded two residues $i$ and $j$
given by $r_{ij}=|\vec{r}_i-\vec{r}_j|$.
If both residues $i$ and $j$ are hydrophobic, $C_{ij}=D_{ij}=1$.
If one residue is hydrophilic and the other hydrophilic or hydrophobic,
$C_{ij}=2/3$ and $D_{ij}=-1$.
If one residue is neutral and the other neutral, hydrophilic or hydrophobic,
$C_{ij}=1$ and $D_{ij}=0$.
For convenience, we set $a=\epsilon=\sigma=1$.


\section{conformational space annealing}

The CSA unifies the essential ingredients of the three global optimization methods,
SA, GA and MCM.
First, as in MCM, we consider only the
phase space of local minima, that is, all conformations are energy-minimized by a
local minimizer.
Secondly, as in GA, we consider many conformations
(called a {\it bank} in CSA) collectively,
and we perturb a subset of bank conformations ({\it seeds})
using the information in other bank conformations.
This procedure is similar to mating in GA.
However, in contrast to the mating procedure in GA,
we replace typically {\it small} portions of a seed
with the corresponding parts of bank conformations
since we want to search the neighborhood of the seed conformation.
Finally, as in SA, we introduce an annealing parameter $D_{\rm cut}$
(a cutoff distance in the phase space of local minima),
which plays the role of temperature in SA.
The diversity of sampling is directly controlled in CSA
by introducing a distance measure between two conformations
and comparing it with $D_{\rm cut}$,
whereas in SA there are no such systematic controls.
The value of $D_{\rm cut}$ is slowly reduced just as in SA,
hence the name {\it conformational space annealing}.
Maintaining the diversity of the population using a distance measure was
also tried in the context of GA, although no annealing was performed \cite{hartke}.
To apply the CSA to an optimization problem,
only two things are necessary;
a method for perturbing a seed conformation,
and a distance measure between two conformations.
This suggests that the CSA is a candidate for a versatile
and yet powerful global optimization method.


The way we picture the phase space of local minima is as follows (see Figure 1).
We assume that most of the phase space of local minima
can be covered by a finite number of large spheres with radius $D_{\rm cut}$,
which are centered on randomly chosen minima ({\it bank}).
Each of the bank conformations is supposed to represent
all local minima contained in
the sphere centered on it.
To improve a bank conformation $A$,
we first select $A$ as a seed.
We perturb $A$ and subsequently energy-minimize it to generate
a trial conformation $\alpha$.
Since $\alpha$ originates from $A$ by small perturbation,
it is likely that $\alpha$ is
contained in a sphere centered on $A$.
If the energy of $\alpha$ is
lower than that of $A$, $\alpha$ replaces $A$ and the center of
the sphere moves from $A$ to $\alpha$.
If it happens that $\alpha$ belongs to a different sphere centered on $B$,
$\alpha$ can replace $B$ in a similar manner.
When $\alpha$ is outside of all existing spheres,
a new sphere centered on $\alpha$ is generated.
In this case, to keep the total number of
spheres fixed, we remove the sphere represented by the highest-energy
conformation.
Obviously, the former two cases are more likely to happen
when the spheres are large, and the latter when spheres are small.
Consequently, larger value of $D_{\rm cut}$ produces more diverse sampling,
whereas smaller value results in quicker search of low-energy conformations
at the expense of getting trapped in
a basin probably far away from the global minimum.
Therefore, for efficient sampling of the phase space, it is necessary to
maintain the diversity of sampling in the early stages and then
gradually shift the emphasis toward obtaining low energy conformations,
which is realized, in CSA, by slowly reducing the value of $D_{\rm cut}$.

When the energy of a seed conformation does not improve
after a fixed number of perturbations,
we stop perturbing it.
To validate this judgment,
it is important that typical perturbations are kept small,
so that the perturbed conformations are close to their
original seeds.
When all of the bank conformations are used as seeds
(one iteration completed), this implies
that the procedure of updating the bank might have
reached a deadlock.
If this happens we reset all bank conformations to be eligible
for seeds again, and we repeat another iteration.
After a preset number of iterations,
we conclude that our procedure has reached a deadlock.
When this happens,
we enlarge the search space by adding more random conformations
into the bank and repeat the whole procedure
until the stopping criterion is met.



In the application of the CSA method to the BLN model protein (see Figure 2),
we first randomly generate a certain number of initial
conformations (for example, fifty random conformations) whose energies are subsequently
minimized using Gay's secant unconstrained minimization solver, SUMSL \cite{gay}.
In the following of the article the term {\it minimization} is used to
refer to the application of SUMSL to a given conformation.
We call the set of these conformations the {\it first bank}.
We make a copy of the first bank and call it the {\it bank}.
The conformations in the bank are updated in later stages,
whereas those in the first bank are kept unchanged.
Also, the number of conformations in the bank is kept unchanged
when the bank is updated. The initial value of $D_{\rm cut}$ is
set as $D_{\rm ave}/2$ where $D_{\rm ave}$ is the average distance
between the conformations in the first bank.
New conformations are generated by choosing a certain number
of {\it seed} conformations (for example, ten seed conformations)
from the bank and by replacing parts of the seeds
by the corresponding parts of conformations randomly chosen
from either the first bank or the bank.
For example, five conformations are generated
for each seed using the partial replacements.
Then the energies of these conformations are subsequently minimized,
and these minimized conformations become trial conformations.

A newly obtained local minimum conformation $\alpha$ is compared with those
in the bank to decide how the bank
should be updated. One first finds the conformation $A$ in the bank
which is closest to the trial conformation $\alpha$ with the distance $D(\alpha, A)$
defined by
\bea
D(\alpha,A)
&=&\sum_{i=1}^{M-3}\min[{\rm mod}(|\phi_i^\alpha-\phi_i^A|,360^\circ),\cr
& &360^\circ-{\rm mod}(|\phi_i^\alpha-\phi_i^A|,360^\circ)],
\eea
where ${\rm mod}(A,B)$ is the least positive value of $x$ satisfying $A=nB+x$
with an integer $n$.
If $D(\alpha, A) < D_{\rm cut}$, $\alpha$ is considered as similar to $A$.
In this case, the conformation with lower energy among $\alpha$ and $A$ is
kept in the bank, and the other one is discarded.
However, if $D(\alpha, A) > D_{\rm cut}$, $\alpha$ is regarded as distinct
from all conformations in the bank. In this case,
the conformation with the highest energy among the bank conformations
plus $\alpha$ is discarded, and the rest are kept in the bank.
We perform this operation for all trial conformations.

After the bank is updated, the $D_{\rm cut}$ is reduced by a fixed ratio,
in such a way that $D_{\rm cut}$ reaches
$D_{\rm ave}/5$ after $L$ local minimizations (for example, $L=500$).
Then seeds are selected again from the bank conformations
which have not been used as seeds yet, to repeat aforementioned procedure.
The value of $D_{\rm cut}$ is kept constant after it reaches the final value.
When all conformations in the bank are used as seeds,
one round of iteration is completed.
We perform an additional search by erasing the record of bank conformations
having been used as seeds, and starting a new round of iteration.
After three iterations are completed,
we increase the number of bank conformations by adding
fifty randomly generated and minimized conformations into the bank
(and also into the first bank),
and reset $D_{\rm cut}$ to $D_{\rm ave}/2$.
The algorithm stops when the known global minimum is found,
which is examined after the bank is updated by all trial conformations.
It should be noted that since one iteration is completed
only after all bank conformations have been used as seeds,
and we add random conformations whenever our search has reached a deadlock,
there is no loss of generality for using particular values for
the number of seeds, the number of bank conformations, etc.


\section{results and discussion}


We applied the CSA method to the BLN 46-mer
with the sequence $B_9N_3(LB)_4N_3B_9N_3(LB)_5L$.
We carried out 100 independent runs and found the global minimum-energy
conformation with energy
\be
E_0=-49.2635
\ee
for all 100 runs.
Figure 3 shows a scatter plot of the number of function evaluations
to obtain the global minimum-energy conformation for 100 independent runs.
On average, it took about $1.865\times10^6$ function evaluations for each run.
The total running time for all 100 runs was 48 hours and 8 minutes
on an Athlon processor (1.8 GHz).
For a single run it took only about 29 minutes on average
to obtain the global minimum-energy conformation.
For the 46-mer, it is reported \cite{ylee00a,ylee01} that the average success rate
to reach the global minimum
is about 9\% using SA with $32\times10^6$ Monte Carlo sweeps
(1 Monte Carlo sweep = 46 Monte Carlo steps).
For the purpose of fair benchmarking, we also carried out hundreds of the SA runs and
we found that our results are consistent with these numbers.
A single run of SA with $32\times10^6$ Monte Carlo sweeps took about 34 hours on the same Athlon processor.
Therefore, the CPU time for CSA is about seventy times less than that for SA,
and its success rate (100\%) is about eleven times higher.
Until now, the best method to obtain the global minimum of the 46-mer
has been the quantum thermal annealing with renormalization (QTAR) \cite{ylee01}.
The QTAR method has been the only method with 100\% success rate
and yet requires 7.5 times less computation than SA.
Therefore, compared to QTAR, CSA is about ten times more efficient
in finding the global minimum of the BLN 46-mer.



One of the attractive aspects of CSA is that a population of distinct
local minima is obtained as a by-product, in addition to the global
minimum-energy conformation.
Gathering all conformations in the final banks of 100 independent CSA runs,
we have obtained a total of 6050 conformations all together.
The conformation with the highest energy among the 6050 conformations has the energy
\be
E_h=-36.355.
\ee
Table I shows the energy spectrum of these 6050 conformations.
There are 2361 different energy levels between $E_0$ and $E_h$.
As the energy level becomes higher, the number of energy levels increases fast.
This implies that conformations near the global minimum are
discretely distributed but conformations at higher energies
are quasi-continuum states.
In Table I the number of energy levels decreases for $E>-43$
due to the fact that conformations
of high energies are discarded as a CSA run proceeds.
Figure 4 shows the number of conformations as a function of energy for $E\le-48$.
It should be noted that there exist two largest energy gaps
\be
\Delta E_a=0.201
\ee
between $E_5=-48.939$ and $E_6=-48.738$
and
\be
\Delta E_b=0.138
\ee
between $E_8=-48.731$ and $E_9=-48.593$.



Since the potential energy of BLN model proteins is invariant
under the inversion transformation
\be
\vec{r_i}\to-\vec{r_i}\ \ \ {\rm or}\ \ \ \phi_i\to-\phi_i,
\ee
two conformational states exist for each energy level
(two-fold degeneracy), including the global minimum-energy conformation.
The root-mean-square deviation (RMSD) between the two global minima
is 0.862.
The inversion symmetry is employed in CSA by generating conformations
with positive values only for the first dihedral angle $\phi_1$ ($0\le\phi_1\le180$).
Figure 5 shows the RMSD values as a function of energy for all
2361 conformations with $0\le\phi_1\le180$, calculated from the global minimum.
In Figure 5 most of the 2361 conformations have small values of RMSD
from the global minimum, independent of their diverse energy values.
These conformations are grouped together in the region RMSD $<1.6$,
separated from conformations with larger values of RMSD.
The number of conformations with RMSD values less than 1.6
is 2215, which is 93.8\% of all conformations.
It is quite interesting to observe that no conformations
with energy less than $-44.34$ appear in the region RMSD $>1.6$.
We also performed 100 CSA runs without employing the inversion symmetry.
The results are quite similar to those in Table I and Figures 3, 4, and 5,
and finding the global minimum requires about 1.4 times more computation.

Now we examine the RMSD-versus-energy distribution for conformations
near the global-minimum energy $E_0$ in detail.
Figure 6 is an enlargement of the lower left corner of Figure 5
for $E\le-48$.
Similarly, Figure 7 shows the results with $-180\le\phi_1\le180$.
The values of RMSD are calculated from one of the two global minima.
The RMSD distribution calculated from the other global minimum is similar to Figure 7.
The twelve (six pairs) conformations below the largest gap $\Delta E_a$ (Eq.~(9))
in Figure 7 are of two branches,
which are related to each other by the inversion transformation.
We call ten of these conformations $(E_1=-49.186$, $E_2=-49.149$, $E_3=-49.063$,
$E_4=-49.002$, and $E_5=-48.939$) as quasi-gobal-minima,
since they are more frequently obtained as final candidates for global minimum,
in SA, than the native state is.
Once a SA run is trapped into one of these quasi-global-minima,
the trapped conformation cannot be folded into the global minimum before crossing over
the two large energy barriers of Eqs.~(9) and (10).
This clearly demonstrates the reason for the poor success rate of SA.
On the other hand, the RMSD values for the conformations above the energy gap
are scattered between 0.15 and 1.2.
In particular, conformations with energy $E_8=-48.731$ (RMSD $=0.156$, 0.881)
can be easily folded into the native state
without being trapped in one of the quasi-global-minima.


\section{conclusion}

We have applied the versatile and powerful global optimization method,
conformational space annealing, to the 46-residue BLN protein with a sequence
$B_9N_3(LB)_4N_3B_9N_3(LB)_5L$.
The CSA method always maintains the diversity of sampling
and is able to cross the high energy barriers between local minima.
Consequently, this method not only finds the global minimum-energy conformation
successfully and efficiently
but also investigates many other distinct local minima as a by-product.

This 46-residue BLN protein is folded into the native state of
a four-stranded $\beta$-barrel.
It has been a challenging problem to locate the global minimum of the BLN 46-mer
because the molecule is highly frustrated and its energy landscape is quite rugged.
We conducted 100 independent CSA runs and successfully found the global
minimum-energy conformation of the 46-residue BLN protein for all 100 runs.
The CPU time for CSA is about seventy times less than that for simulated annealing,
and its success rate (100\%) to find the global minimum is about eleven times higher.
In addition, the amount of computational efforts required for CSA
to find the global minimum is about ten times less than
that for the QTAR, the best global optimization method
yet applied for the 46-residue BLN protein.

The 100 separate CSA runs produce the global minimum 100 times as well as
other 5950 final conformations corresponding to 2360 distinct local minima
of the 46-residue BLN protein.
As the values of energy levels becomes higher,
the number of energy levels (or the number of local minima) increases fast.
Conformations near the global minimum are discretely distributed but
conformations of higher energies are quasi-continuum states.
Most of final conformations have relatively small RMSD values from the global
minimum, independent of their diverse energy values.
Very close to the global minimum, there exist quasi-global-minima which
are frequently obtained as one of the final answers from SA runs.
We find that there exist two largest energy gaps between the quasi-global-minima
and the other local minima.
Once a SA run is trapped in one of these quasi-global-minima,
it cannot be folded into the global minimum before crossing over
the two large energy barriers,
clearly demonstrating the reason for the poor success rate of SA.


\begin{acknowledgments}
We are grateful to Prof. Julian Lee for useful discussions
and for kindly drawing Figure 1.
\end{acknowledgments}



\newpage

\begin{table}
\caption{The energy spectrum of 6050 final conformations
for all 100 independent runs of CSA.
NE is the number of energy levels for a given energy range,
NC is the number of conformations for a given energy range,
and DC is the density of conformations
(the number of conformations per energy level, that is, DC = NC/NE).
The total number of energy levels is 2361.}
\begin{ruledtabular}
\begin{tabular}{crrr}
energy range &NE &NC &DC \\
\hline
$E_0\le E\le-49$ &5 &295 &59.00 \\
$-49<E\le-48$ &31  &722  &23.29 \\
$-48<E\le-47$ &111 &1033 &9.31 \\
$-47<E\le-46$ &192 &699  &3.64 \\
$-46<E\le-45$ &297 &689  &2.32 \\
$-45<E\le-44$ &374 &766  &2.05 \\
$-44<E\le-43$ &377 &624  &1.66 \\
$-43<E\le-42$ &353 &517  &1.46 \\
$-42<E\le-41$ &263 &319  &1.21 \\
$-41<E\le-40$ &159 &181  &1.14 \\
$-40<E\le-39$ &110 &114  &1.04 \\
$-39<E\le-38$ &50  &51   &1.02 \\
$-38<E\le-37$ &27  &28   &1.04 \\
$-37<E\le E_h$ &12 &12   &1.00 \\
\end{tabular}
\end{ruledtabular}
\end{table}


\begin{figure}
\includegraphics[width=13cm]{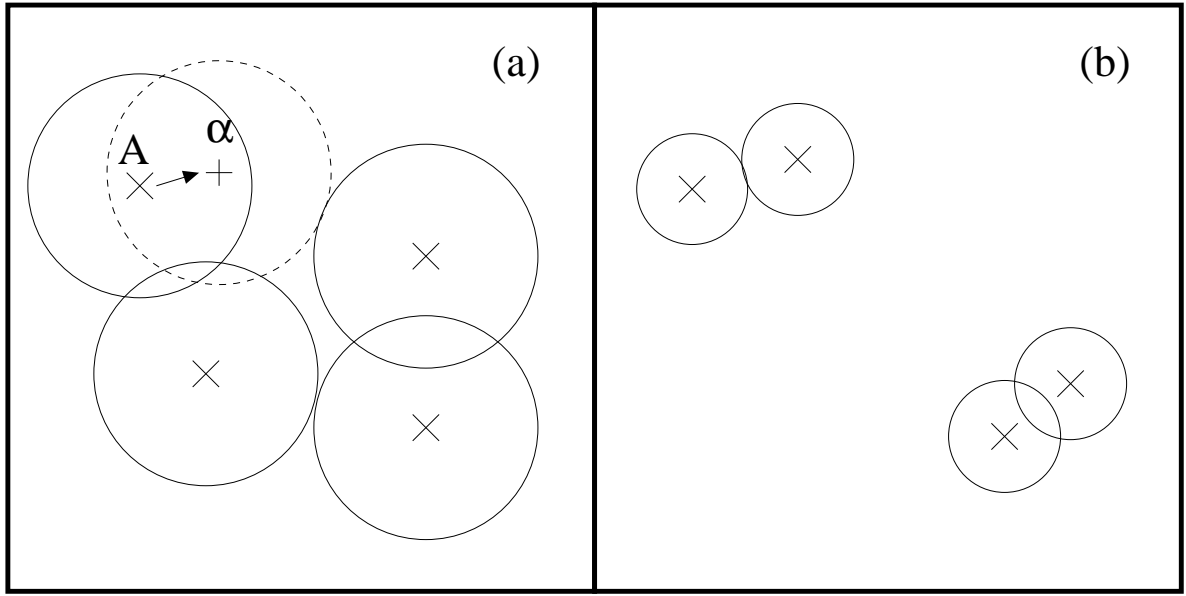}
\caption{A schematic for the search procedure of CSA.
The boxes represent the identical phase space.
(a) Initially, we cover the phase space by large spheres centered
on randomly chosen local minima denoted by $\times$ symbols,
and replace the centers with lower-energy local minima.
When an initial conformation $A$ is replaced by a new conformation $\alpha$,
the sphere moves in the direction of the arrow.
(b) As the CSA algorithm proceeds and the energies of the representative conformations
at the centers of the spheres are lowered, the size of the spheres are reduced
and the search space is narrowed down to small basins of low-lying local minima.}
\end{figure}

\begin{figure}
\includegraphics{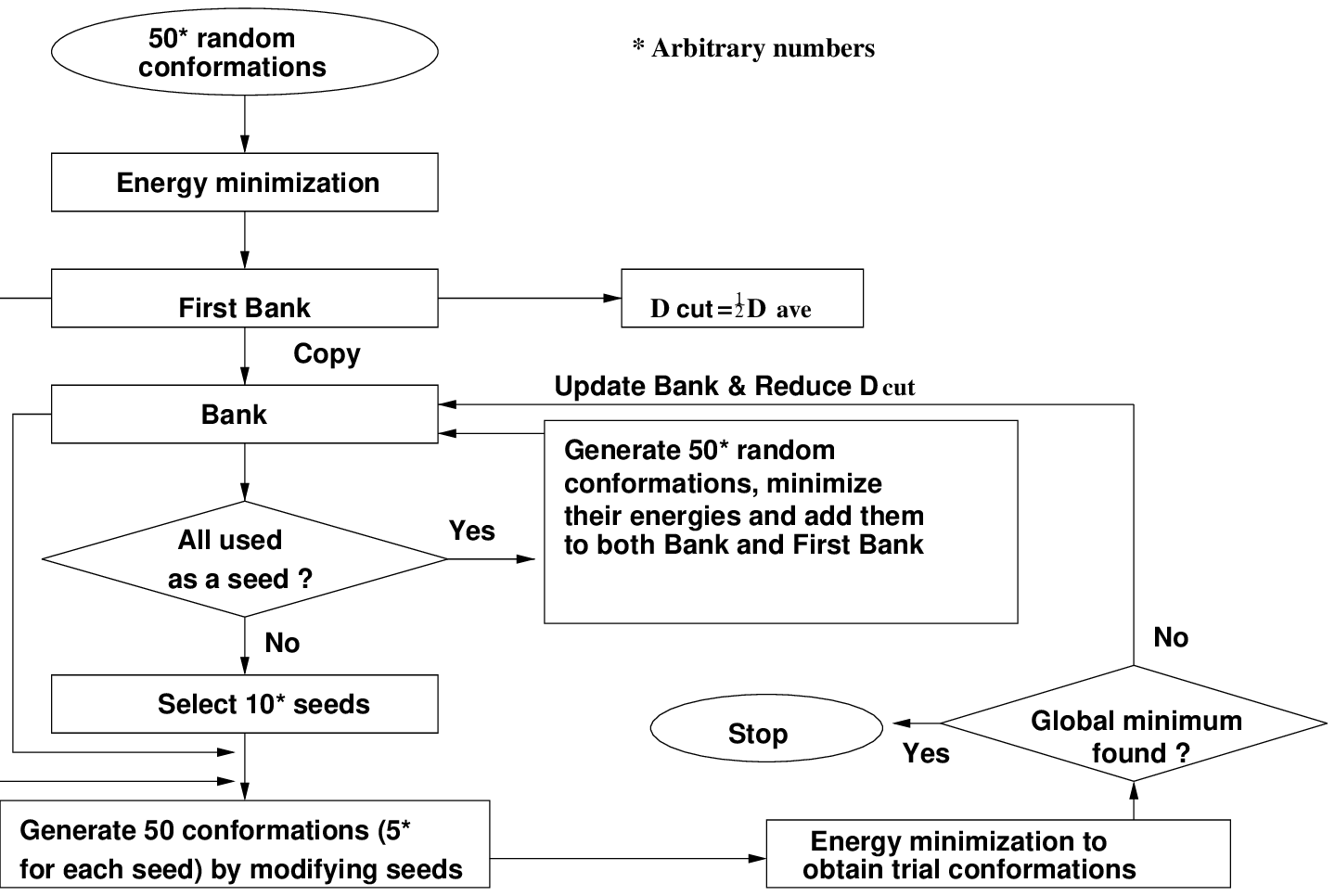}
\caption{Flow chart of the CSA algorithm to locate the global minimum
of a BLN model protein.}
\end{figure}

\begin{figure}
\includegraphics[width=16cm]{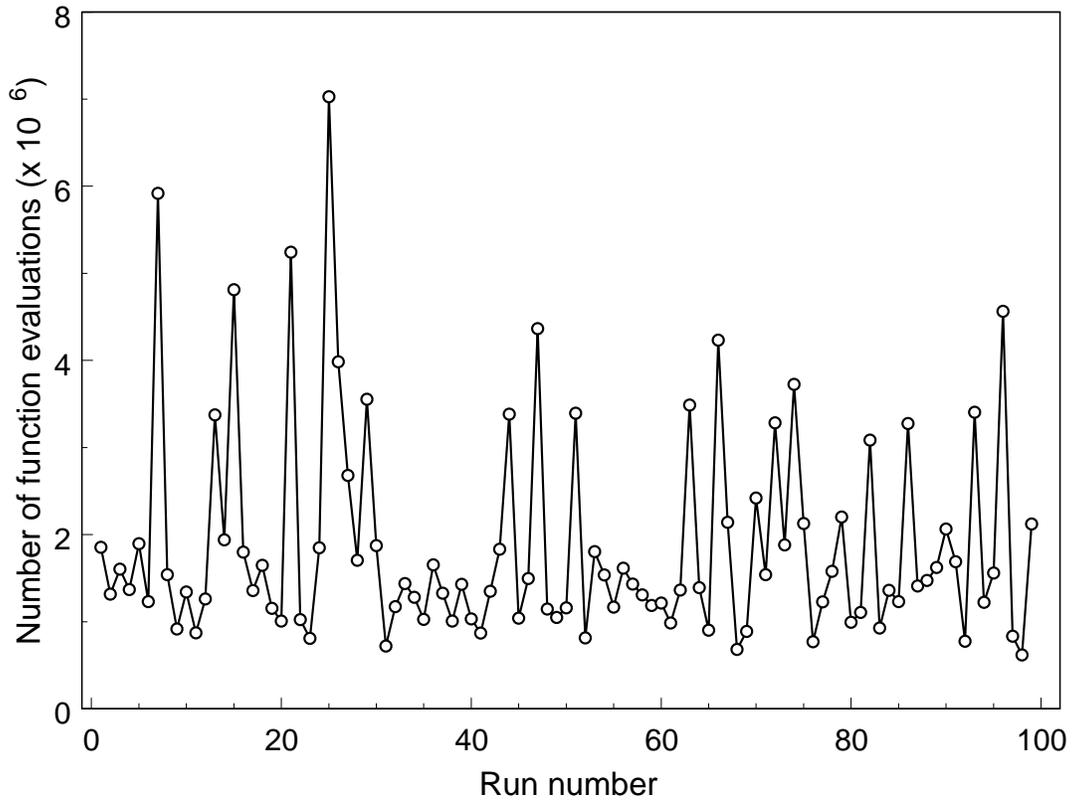}
\caption{Scatter plot of the number of function evaluations to obtain
the global minimum-energy conformation with energy $E_0=-49.2635$ for
100 independent runs. The global minimum-energy conformation was successfully
obtained for all 100 independent runs with about
$1.865\times10^6$ function evaluations on average.}
\end{figure}

\begin{figure}
\includegraphics[width=16cm]{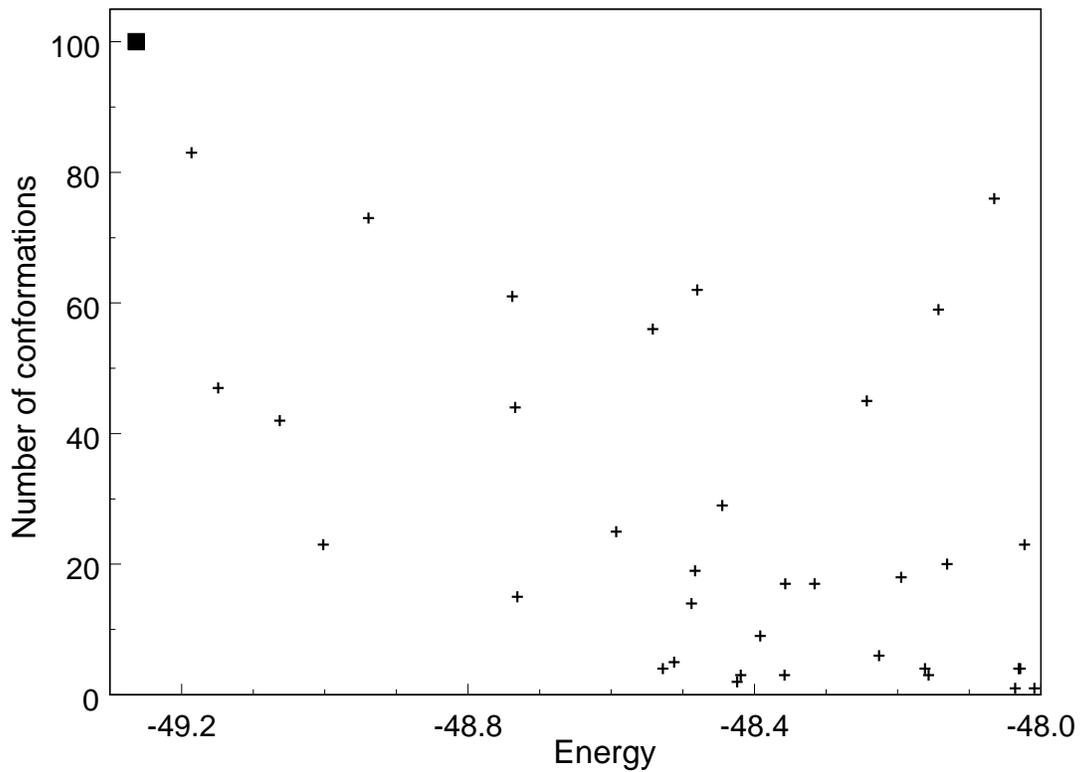}
\caption{The number of conformations as a function of energy for $E\le-48$,
accumulated from all 100 independent runs.
The closed square represents the number of the global minimum-energy conformation
with energy $E_0=-49.2635$.}
\end{figure}

\begin{figure}
\includegraphics[width=16cm]{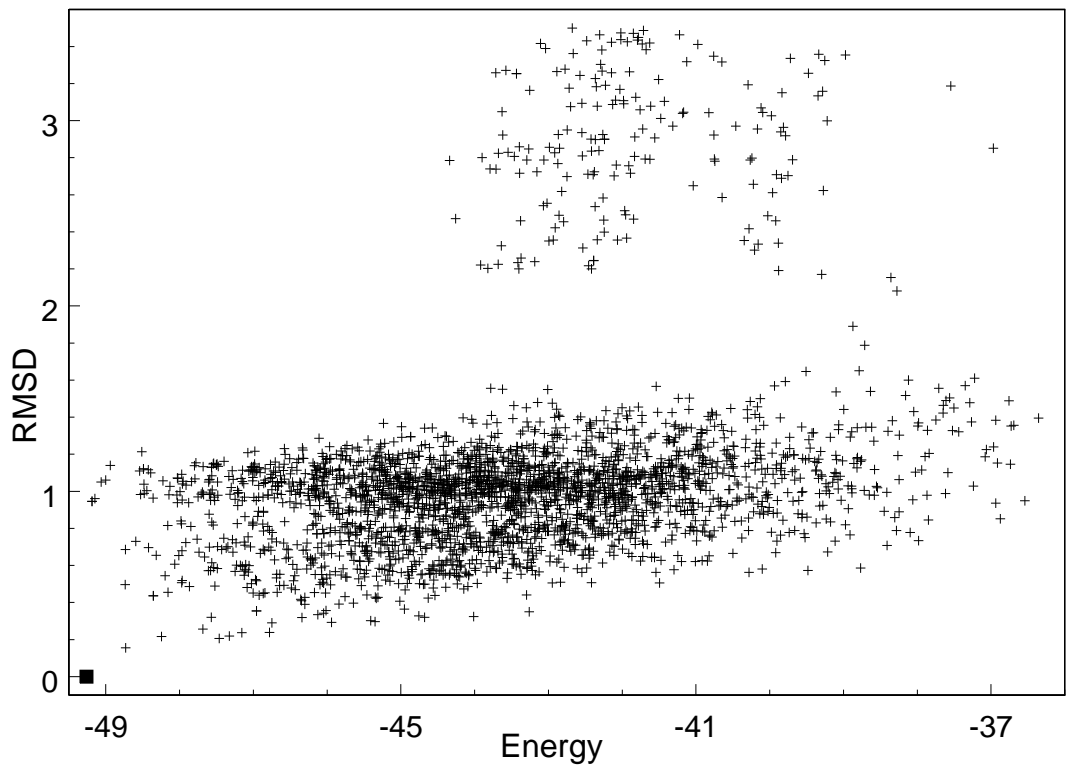}
\caption{Distribution of the RSMD values as a function of energy for all
2361 conformations calculated from one of the two global minima.
The closed square represents the global minimum-energy conformation
with energy $E_0=-49.2635$.}
\end{figure}

\begin{figure}
\includegraphics[width=16cm]{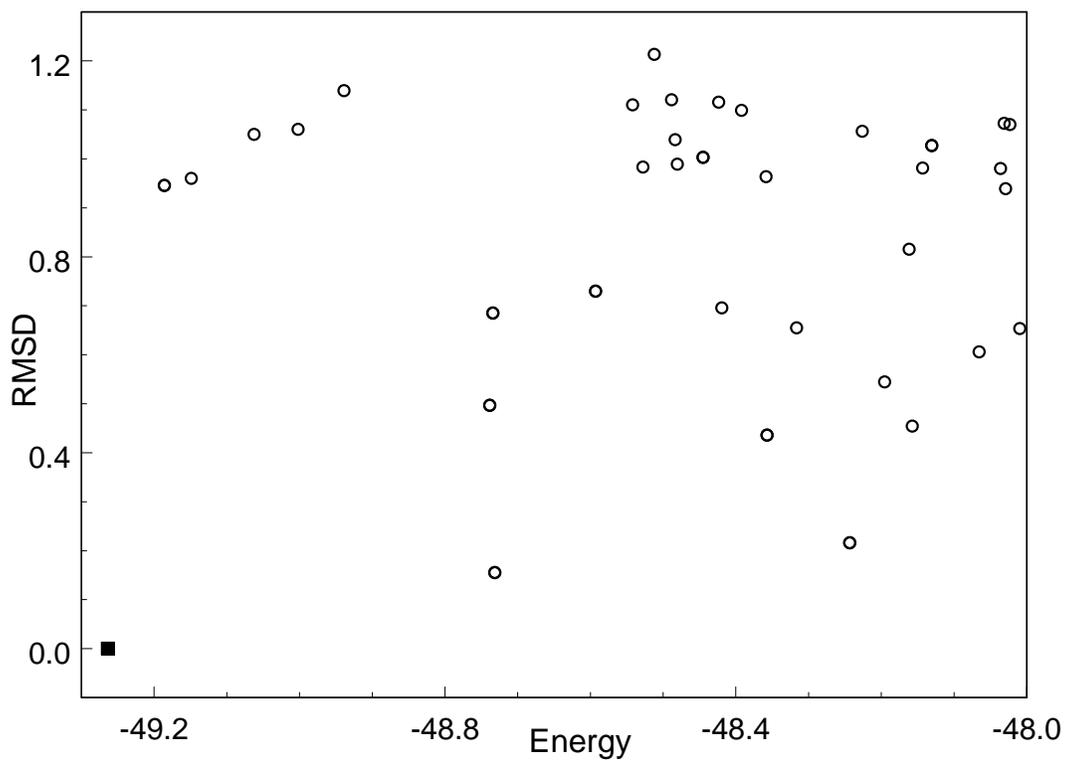}
\caption{Distribution of the RMSD values as a function of energy for conformations
with energy $E\le-48$ (an enlargement of the lower
left corner of Figure 5).
The closed square represents the global minimum-energy conformation
with energy $E_0=-49.2635$.}
\end{figure}

\begin{figure}
\includegraphics[width=16cm]{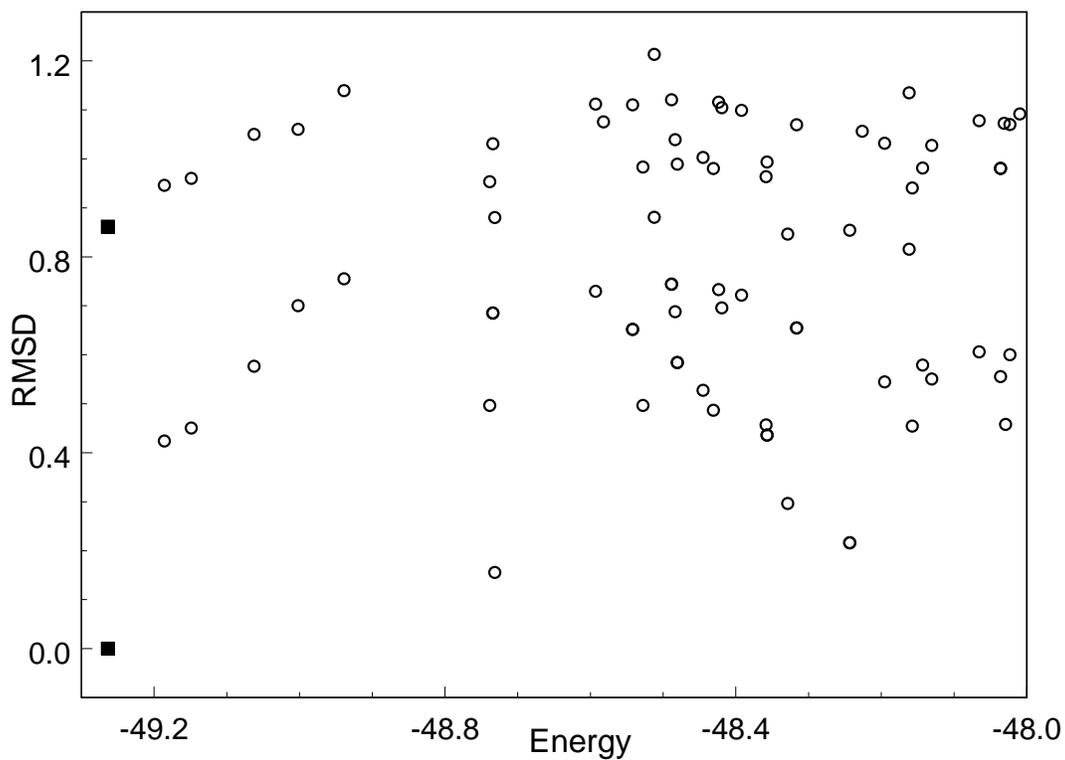}
\caption{Distribution of the RMSD values as a function of energy for conformations
with energy $E\le-48$, obtained from 100 CSA runs without employing
the inversion symmetry.
Two closed squares represent the global minimum-energy conformations
with energy $E_0=-49.2635$ connected by the inversion transformation.}
\end{figure}


\end{document}